\documentclass[aip,apl,amsmath,amssymb,reprint,superscriptaddress]{revtex4-1}
\usepackage{graphicx}
\usepackage{dcolumn}
\usepackage{bm}

\begin{document}
\title{Electron spin ensemble strongly coupled to a three-dimensional microwave cavity}
\author{Eisuke Abe}
\email{eisuke.abe@materials.ox.ac.uk}
\affiliation{Department of Materials, Oxford University, Parks Road, Oxford OX1 3PH, UK}
\author{Hua Wu}
\affiliation{Department of Materials, Oxford University, Parks Road, Oxford OX1 3PH, UK}
\author{Arzhang Ardavan}
\affiliation{CAESR, Clarendon Laboratory, Department of Physics, Oxford University, Oxford OX1 3PU, UK}
\author{John J. L. Morton}
\affiliation{Department of Materials, Oxford University, Parks Road, Oxford OX1 3PH, UK}
\date{\today}

\begin{abstract}
We demonstrate the strong coupling between an electron spin ensemble and a three-dimensional cavity in a reflection geometry.
We also find that an anticrossing in the cavity/spin spectrum can be observed under conditions that
the collective coupling strength $g_c$ is smaller than the spin linewidth $\gamma_s$ or the cavity linewidth.
We identify a ratio of $g_c$ to $\gamma_s$ ($g_c/\gamma_s >$ 0.64) as a condition to observe a splitting in the cavity frequency.
Finally, we confirm that $g_c$ scales with $\sqrt{N}$, where $N$ is the number of polarized spins.
\end{abstract}
\maketitle
When two oscillators couple strongly, their otherwise degenerate modes repel each other to develop an anticrossing in the spectrum.
A canonical example is the interaction between a single photon and a single atom,
and this phenomenon is called the vacuum Rabi splitting.~\cite{WMQO}
The strong light-matter coupling has been a central subject in cavity quantum electrodynamics (QED),
and has attracted increasing attention as a means to coherently transfer information between a flying (photonic) and stationary (atomic) qubits
in the field of quantum information processing.
So far, the strong coupling regime has been realized in various systems.~\cite{BSM+96,BMB+04,RSL+04,WSB+04}
Even when the cavity-atom coupling strength $g_s$ is small, as in the case of natural spins,
the interaction can be collectively enhanced, by utilizing an ensemble of $N$ identical two-level systems,
to replace  $g_s$ with a new parameter $g_c = g_s \sqrt{N}$.
Such a collective enhancement has been observed in a wide variety of systems, such as
Bose-Einstein condensates,~\cite{BDR+07,CSD+07} trapped ions,~\cite{HDM+09} and circuit QED with superconductors.~\cite{FBB+09}

More recently, the strong coupling between an electron spin ensemble and a superconducting coplanar waveguide resonator
has been demonstrated for impurity spins in ruby and diamond.~\cite{SSG+10,KOB+10}
A rough estimation of $g_s$ in the spin system is given as $g_s \approx m_0 \sqrt{\mu_0 \omega_c / 2\hbar V_c}$,
where $m_0$ is the magnetic dipole moment of the spin,
$\omega_c$ is the cavity frequency, and $V_c$ is the cavity mode volume.
There, the small $V_c$ of the order of 10$^{-12}$ m$^3$ helped to enhance the cavity-spin coupling.
These experiments are motivated by the goal of using an electron spin ensemble as a medium for quantum memory,
which will potentially benefit from the capability of storing multiple qubits in different modes of the ensemble.~\cite{WGW+10}

In this Letter, we study the coupling between an electron spin ensemble and a microwave resonator with much larger $V_c$,
a three-dimensional cavity that is commonly used for bulk electron paramagnetic resonance (EPR),
and demonstrate that by increasing the number of spins the spectrum experiences the transition from the weak to strong coupling regimes.
Our result provides an opportunity for studying the strong coupling effect
with readily accessible setups and wider tunabilities in the experimental parameters.
In addition, it has been recently observed that
such a three-dimensional cavity allows a superconducting qubit to preserve its coherence longer than
the case of a two-dimensional counterpart.~\cite{PSB+11}
Together with this observation, our result makes a three-dimensional cavity a promising component
for hybrid-qubit architectures.~\cite{TML03,ADD+06,I09,WAB+09}
We note that Chiorescu {\it et al.} have made a similar observation in an electron spin ensemble utilizing a cavity ringing effect,~\cite{CGB+10}
but did not address the $\sqrt{N}$-dependence of $g_c$ nor the parameter regime we report here.

The microwave cavity we used is an X-band cylindrical dielectric ring resonator from Bruker (ER 4118X-MD5).
$V_c$ of this resonator is about 2 $\times$ 10$^{-7}$ m$^3$, giving an expected $g_s/2\pi$ of 0.06~Hz.
The reflection spectrum ($S_{11}$) from the resonator was measured with a network analyzer (HP 8722D).
As will be discussed later, in all the measurements, the sample position inside the cavity was carefully adjusted to maximize the coupling.

Two materials were examined in this study;
the first material is 2,2-diphenyl-1-picrylhydrazyl (DPPH) powder,
a radical commonly used for field calibration of EPR spectrometers.
It has an EPR linewidth of 2.1~G at room temperature.
In this material, the spin concentration can be made high;
our samples had a total number of spins $N_{\mathrm{tot}}$
ranging from 1.0 $\times$ 10$^{18}$ to 2.4 $\times$ 10$^{19}$.
As the spins are in thermal equilibrium at room temperature,
the number of polarized spins, which gives $N$ in the present study,
is given as $N_{\mathrm{tot}} \times (h~9.8~\mathrm{GHz}/2 k_B~300~\mathrm{K})$,
and varies from 7.8 $\times$ 10$^{14}$ to 1.9 $\times$ 10$^{16}$.
The second material is x-form lithium phthalocyanine (LiPc) powder electrochemically prepared from Li$_2$Pc.~\cite{IZK00}
By reducing the partial pressure of oxygen, this radical has a very narrow EPR linewidth, in our case around 83~mG at room temperature.
On the other hand, $N$ in these samples was much smaller than that of DPPH:
less than 7.4 $\times$ 10$^{14}$ (including the polarization factor).
The number of spins was determined with an X-band cw EPR spectrometer (Bruker EMX), independently from the measurements described below.

\begin{figure}
\begin{center}
\includegraphics{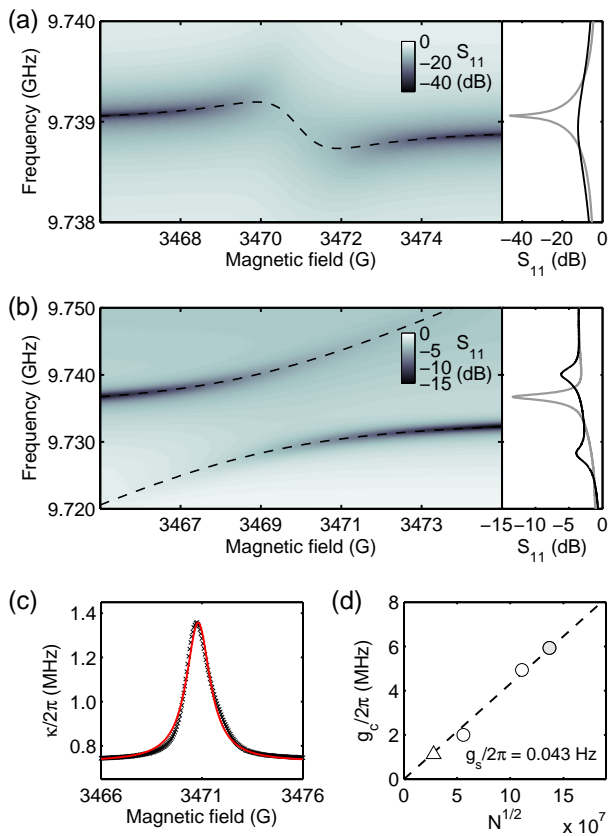}
\caption{(Color online)
(a) Reflection spectrum of DPPH with $N$ = 7.8 $\times$ 10$^{14}$ in a logarithmic scale.
The input microwave power was $-$20~dBm.
The dashed line is a fit by Eq.~(\ref{ofit}).
The right panel shows the cross sections at 3470.9~G (black, on resonance) and 3466~G (gray).
(b) Reflection spectrum of DPPH with $N$ = 1.9 $\times$ 10$^{16}$.
The dashed lines are fits by Eq.~(\ref{vrs}), giving the splitting of $2g_c/2\pi$ = 11.9~MHz.
The cross sections at 3469~G (black, on resonance) and 3465~G (gray) are given in the right panel.
(c) Change of $\kappa$ extracted from (a). The solid line is a Lorentzian fit.
(d) $g_c/2\pi$ of DPPH as a function of $\sqrt{N}$.
The triangular point ($\bigtriangleup$) is obtained from the analysis in (c),
whereas circular points ($\bigcirc$) are obtained from fits using Eq.~(\ref{vrs}).
The point in gray corresponds to (b).
}
\label{dpph}
\end{center}
\end{figure}

We first discuss the result from DPPH.
Figure~\ref{dpph}-(a) shows the reflection spectrum for the $N$ = 7.8 $\times$ 10$^{14}$ sample.
Here we observe features familiar in conventional EPR.
As is well-known, the dispersion and absorption components of the EPR signals
are mathematically connected each other by the Kramers-Kronig relations.~\cite{PooleESR}
The former appears as the shift of the cavity frequency across the resonance (at $B_r$ = 3470.9~G),~\cite{afc}
while the change in the cavity linewidth, or cavity quality factor $Q$, is a consequence of the latter.
A model with two coupled oscillators describes the changes in cavity frequency $\omega$ and cavity half-width $\kappa$ as~\cite{HDM+09}
\begin{eqnarray}
\omega = \omega_c -  g_c^2 \Delta/(\Delta^2 + \gamma_s^2), \label{ofit}\\
\kappa = \kappa_c + g_c^2 \gamma_s/(\Delta^2 + \gamma_s^2) \label{kfit},
\end{eqnarray}
where $\kappa_c$ is the cavity half-width far from the resonance,
$\gamma_s$ is the spin half-width, and $\Delta = m_0 (B - B_r)/\hbar$ is the the field detuning.
A fit to the cavity dip positions in Fig.~\ref{dpph}-(a) gives $g_c/2\pi$ = 1.15~MHz and $\gamma_s/2\pi$ = 2.85~MHz.
Figure~\ref{dpph}-(c) plots $\kappa$ as a function of external magnetic fields.
A fit using Eq.~(\ref{kfit}) gives $g_c/2 \pi$ = 1.12~MHz, $\gamma_s/2\pi$ = 2.00~MHz, and $\kappa_c/2\pi$ = 0.73~MHz.
The values of $g_c$ obtained from the two fits agree well as expected.
Those of $\gamma_s$ slightly differ from each other but are consistent with that from cw EPR,
in which the Gaussian full-width of 2.1~G corresponds $\gamma_s/2\pi$ = 3.5~MHz.

As we increase $N$, the spectrum changes dramatically.
When $N$ is increased to four times that of the previous measurement, we start to observe two clearly separated dips,
and the splitting becomes larger as we increase $N$.
Figure~\ref{dpph}-(b) shows the case for $N$ = 1.9 $\times$ 10$^{16}$.
The upper and lower branches $\omega_{\pm}$ are fitted by the expression for the vacuum Rabi splitting
\begin{equation}
\omega_{\pm} = \omega_c + \frac{\Delta}{2} \pm \frac{\sqrt{\Delta^2 + 4 g_c^2}}{2},
\label{vrs}
\end{equation}
from which we deduce $g_c/2\pi$ = 5.9~MHz.
Figure~\ref{dpph}-(d) shows the $\sqrt{N}$-dependence of $g_c$.
The circular points represent samples in which the splitting was evident and from the fit we obtain $g_s/2\pi$ = 0.043~Hz,
which is in reasonable agreement with our estimation of 0.06~Hz.
The triangular point is for the $N$ = 7.8 $\times$ 10$^{14}$ sample that did not show splitting,
but it is apparent that the value of $g_c$ extracted from the change in $Q$-factor also falls onto the fitted line to obey the $\sqrt{N}$-dependence.

In theory, the observation of vacuum Rabi splitting
and the realization of the strong coupling regime
require the condition $g_c \gg \kappa_c, \gamma_s$ or,
similarly, the cooperativity $C = g_c^2/2 \kappa_c \gamma_s \gg 1$.
This is barely satisfied in the $N$ = 1.2 $\times$ 10$^{16}$ and 1.9 $\times$ 10$^{16}$ samples,
where $g_c/2\pi$ = 4.9~MHz and 5.9~MHz are larger than both $\gamma_s/2\pi$ = 3.5~MHz and $\kappa_c/2\pi <$ 1~MHz
(giving $C \approx$ 4 and 7, respectively).
On the other hand the $N$ = 7.8 $\times$ 10$^{14}$ sample has $g_c$ larger than $\kappa_c$ but smaller than $\gamma_s$,
and $C$ of 0.2, consistent with the absence of the anticrossing.
A question arises for the $N$ = 3.1 $\times$ 10$^{15}$ sample, where $g_c/2\pi$ = 2.0~MHz is comparable with, but smaller than
$\gamma_s/2\pi$ ($C$ = 0.8) and yet we observed the splitting.
This observation motivates us to explore a different parameter regime where $g_c$ is much smaller than either $\kappa_c$ or $\gamma_s$.
This is possible with LiPc, in which, according to $N$ available, $g_c/2\pi$ will be less than 1~MHz
while $\kappa_c/2\pi$ can be larger than 1~MHz since this material is more lossy than DPPH.
Owing to its narrow EPR linewidth, $\gamma_s/2\pi$ is only 0.14~MHz, which may compensate the reductions in the other two parameters.

\begin{figure}
\begin{center}
\includegraphics{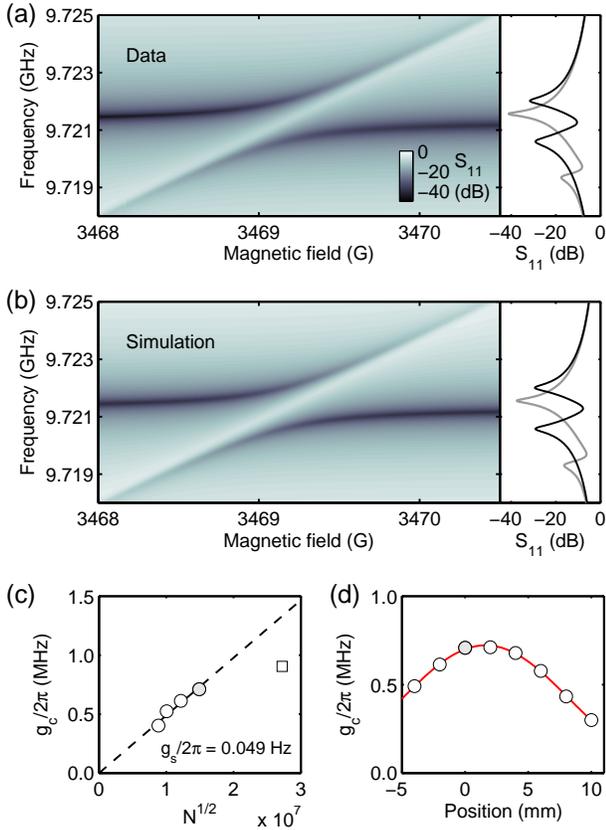}
\caption{(Color online)
(a) Reflection spectrum of LiPc with $N$ = 2.2 $\times$ 10$^{14}$.
The cross sections are given for 3469.2~G (black, on resonance) and 3468.5~G (gray).
(b) Simulation to (a).
The same color bar as (a) applies.
See the text for detail.
(c) $g_c/2\pi$ of LiPc as a function of $\sqrt{N}$.
The square point ($\Box$) is omitted from the fit.
(d) $g_c/2\pi$ of the $N$ = 2.2 $\times$ 10$^{14}$ sample as a function of the sample position along the long axis of the cylindrical resonator.
The origin of the horizontal axis is approximately the center of the resonator.
The solid line is a fit with a sinusoidal function.
In (c) and (d) the point in gray corresponds to (a).
}
\label{lipc}
\end{center}
\end{figure}

Figure~\ref{lipc}-(a) demonstrates an anticrossing behavior in LiPc,
where $g_c/2 \pi$ = 0.71~MHz and $\kappa_c/2\pi$ = 5.4~MHz are experimentally obtained.
The appearance of two dips is obvious, although even at the resonance they are not fully separated (note that the data is given in dB units).
As shown in Fig.~\ref{lipc}-(c), the $\sqrt{N}$-dependence of $g_c$ is confirmed again with nearly the same $g_c$ as in the case of DPPH
(0.043~Hz for DPPH and 0.049~Hz for LiPc).
As for the $N$ = 7.4 $\times$ 10$^{14}$ sample, which was not used for the fitting, $g_c$ deviates from the $\sqrt{N}$-dependence.
We attribute this as due to its very large sample volume compared with other samples,
and the ac magnetic field the sample experiences is no longer uniform.
We can estimate the distribution of the ac magnetic field by simply moving the sample across the cavity [Fig.~\ref{lipc}-(d)].
Indeed, it was necessary to check the position dependence in all samples to ensure that the coupling was maximized.
Fitting the data with a sinusoidal function and assuming the same distribution in the $N$ = 7.4 $\times$ 10$^{14}$ sample,
we numerically average the coupling strength over the sample length,
and obtain $g_c/2\pi$ = 1~MHz, which is reasonably close to the observed value of 0.9~MHz.

\begin{figure}
\begin{center}
\includegraphics{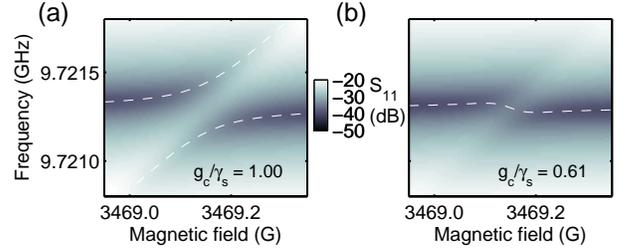}
\caption{(Color online)
(a) Simulation with $g_c/2\pi$ = 0.14~MHz ($g_c/\gamma_s$ = 1).
The dashed lines are Eq.~(\ref{vrs}).
(b) $g_c/2\pi$ = 0.085~MHz ($g_c/\gamma_s$ = 0.61).
The dashed line is Eq.~(\ref{ofit}).
In both cases, all the parameters except for $g_c$ are the same as those for Fig.~\ref{lipc}-(b).
}
\label{sim}
\end{center}
\end{figure}

We now discuss the condition to observe the splitting in more detail,
using the following expression derived from a standard input-output formalism:~\cite{WMQO,SSG+10}
\begin{equation}
|S_{11}|^2 = \left| 1 + \frac{\kappa_e}{i (\omega - \omega_c) - \kappa_c + g_c^2 /(i \Delta - \gamma_s)} \right|^2,
\label{s11}
\end{equation}
where $\kappa_e$ is the external loss.
To demonstrate the validity of Eq.~(\ref{s11}), we compute a spectrum
using the parameter values $g_c/2 \pi$ = 0.71~MHz, $\kappa_c/2\pi$ = 5.4~MHz, $\gamma_s/2\pi$ = 0.14~MHz,
and $\kappa_e$ = 0.99 $\times$ $\kappa_c$ [Fig.~\ref{lipc}-(b)].
This reproduces well the experimental spectrum in Fig.~\ref{lipc}-(a).
Note that in this simulation the only adjustable parameter was $\kappa_e$ and experimentally obtained values were used 
for $g_c$, $\kappa_c$, and $\gamma_s$.
Here, we see $\kappa_c > g_c > \gamma_s$, and $C$ is only 0.3.
The reason we still observe the splitting is understood through a straightforward algebraic analysis of Eq.~(\ref{s11}).
We differentiate it with respect to $\omega$ to find local minima at $\Delta$ = 0.
Assuming $\kappa_e/\kappa_c \approx$ 1, we observe that
two minima exist if $g_c^4 - \gamma_s^2(1 + 4 C)(\gamma_s^2 - 2g_c^2) > 0$ is satisfied.
For $\kappa_c$ large compared with $g_c$ and $\gamma_s$, the condition is simplified to
$g_c/\gamma_s > \sqrt{\sqrt{2} - 1} \approx$ 0.64.
This indicates that the splitting may be observable even if $g_c$ is smaller than both $\gamma_s$ and $\kappa_c$,
as confirmed by additional simulations shown in Fig.~\ref{sim}.
As reducing only $g_c$ while all the other parameters fixed from Fig.~\ref{lipc}-(b),
we observe that two branches merge into one at around $g_c/\gamma_s$ = 0.64.
We emphasize that this does not mean that the light-matter coherent information transfer is possible in this parameter regime.
For the coherent transfer to be possible, the condition $g_c \gg \kappa_c, \gamma_s$ remains a compelling physical requirement.

In conclusion, we have demonstrated the strong coupling between an electron spin ensemble and
a three-dimensional cavity ($g_c > \kappa_c, \gamma_s$) and confirmed the $\sqrt{N}$-dependence of the collective coupling strength $g_c$.
For such a spin ensemble to be useful as a quantum memory, strong coupling and consequently entanglement with a single photon
must be achieved, while the present experiments required many photons (10$^{13}$ or less) to spectroscopically resolve the anticrossing.
In this respect DPPH suffers a relatively large inhomogeneous linewidth, which results in a short storage time
unless multiple microwave pulses for decoupling/refocusing are applied.
We have also observed an anticrossing behavior in the spectrum with the cooperativity $C$ smaller than 1,
and derived the condition for observing the splitting.
An advantage of our system over a cavity QED counterpart lies in the simplicity in the experimental setup;
we may readily choose other spin systems with different EPR linewidths (to control $\gamma_s$) and spin densities (to control $N$),
and $N$ may also be controllable by the temperature or by means of dynamic polarization schemes.
Therefore, we believe that our observation will broaden the applicability of the physics of collective light-matter coupling
to wider parameter regimes.

The authors thank Yasuhiro Ito for help in sample preparation and Andrew Briggs for useful discussions.
AA and JJLM are supported by the Royal Society.


\begin{thebibliography}{9}
\bibitem{afc}
In commercial EPR spectrometers, only the absorption signal is extracted while the dispersion signal is eliminated
by using the automatic frequency control circuitry and locking the source frequency always at the cavity dip position.
\end{thebibliography}
\end{document}